\documentclass{ws-procs9x6}

\begin{document}

\title{Stabilizing Moduli with String Cosmology}

\author{S. Watson}

\address{Department of Physics, \\
Brown University, \\ Providence, RI 02912, USA\\ E-mail:
watson@het.brown.edu}

\maketitle

\abstracts{
In this talk I will discuss the role of finite temperature quantum corrections in string
cosmology and show that they can lead to a stabilization mechanism for the volume moduli.
I will show that from the
higher dimensional perspective this results from the effect
of states of enhanced symmetry on the one-loop free energy.
These states lead not only to stabilization,
but also suggest an alternative model for $\Lambda$CDM.
At late times, when the low energy effective field theory gives the appropriate
description of the dynamics, the moduli will begin to slow-roll and stabilization will generically fail.
However, stabilization
can be recovered by considering cosmological particle production near the points of enhanced symmetry
leading to the process known as moduli trapping.}

\section{String Gases in $ 10D$}
\subsection{Initial Conditions}
One problem in string cosmology is the issue of initial
conditions.  Not only must models of string cosmology address
the standard initial condition problems in cosmology, but string theory
also predicts the existence of extra dimensions.  The usual prescription
for dealing with the extra dimensions is to take them small, stable, and unobservable.
However a complete model of string cosmology should explain how this came about and
why the explicit breaking of Lorentz invariance should be allowed.  A step
in this direction was first proposed by Brandenberger and Vafa\cite{bv}.
They argued that, by
considering the dynamics of a string gas in nine, compact spatial
dimensions initially taken at the string scale, one could
explain why three dimensions grow large while six stay compact.
The crux of their argument is based on the fact that in addition
to the usual Kaluza Klein modes of a particle on a compact space,
strings also possess winding modes.  These extra degrees of freedom
will generically halt cosmological expansion; however, if these
modes could annihilate with their anti-partners this would allow the dimension
they occupy to expand.
Then, the fact that strings generically intersect (interact) in at
most three spatial dimensions means that in the remaining six
dimensions thermal equilibrium cannot be maintained.  Thus,
the winding modes will drop out of equilibrium and the six
spatial dimensions will be frozen near the string scale.
Furthermore, once all the winding modes in the three large dimensions
annihilate, the universe emerges filled with
a gas of momentum modes, which evolves as a
radiation dominated universe.

\subsection{String Gases at Finite Temperature}
The usual starting point of string cosmology is the
action,
\begin{equation}
\label{action1}
S=\frac{1}{2 \kappa^2} \int d^{d+1}x \sqrt{-g} e^{-2 \varphi} \Biggl( R+4 (\nabla \varphi)^2
-\frac{1}{12}H^2 + {\mathcal O}(\alpha^{\prime}) \Biggr) + {\mathcal O}(g_s).
\end{equation}
For simplicity we will ignore the Ramond sector and set $H=0$.
Here we have in
mind the heterotic string on a toroidal background.  Motivated by
the Brandenberger-Vafa scenario we will take the background to be
$\mathbb{R}^4 \times T^6$, where we assume that the three spatial
dimensions have grown large enough to be approximated by an FRW
universe and the six small dimensions are toroidal and near the
string scale.  To include time dependence we make use of the
adiabatic approximation, which implies that we can replace static
quantities by slow varying functions of time.

The action (\ref{action1}) represents a double expansion in both
the string coupling $g_s \sim e^{\langle \varphi \rangle}$ and
the string tension, $T=\frac{1}{4 \pi \alpha^{\prime}}$.  We now want to
include terms coming from $g_s$ corrections at finite
temperature\cite{kr,tv}.  Let us consider the $1$-loop free
energy
\begin{equation}
F=T Z_1 \sim \beta^{-1} \sum e^{-\beta M(n,\omega,N,\bar{N})},
\end{equation}
where $Z_1$ is the one loop partition function,
$M(n,\omega,N,\bar{N})$ is the string mass, and $\beta=1/T$ is the
inverse temperature.  In the early universe we are interested in temperatures
near or below the string scale ($\beta \gtrsim \sqrt{\alpha^{\prime}}$)
where the major contribution to the one-loop free energy can be seen to come from
the massless modes of the string.

\begin{figure}[ht]
\centerline{\epsfxsize=4.0in\epsfbox{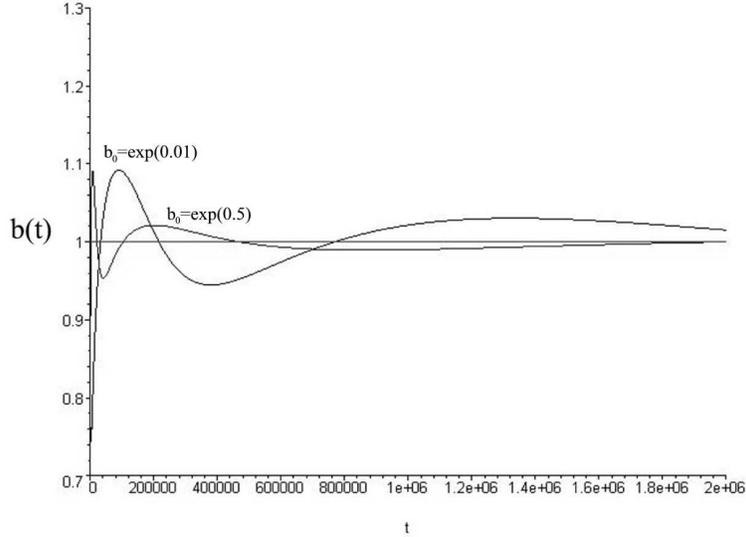}}
\caption{The evolution of the volume modulus of the extra dimensions for different initial values near
the self-dual radius.  We see that the moduli are eventually stabilized at the self-dual radius where the
pressure is found to vanish.\label{fig1}}
\end{figure}

In the case of the heterotic string there are additional massless
states that occur at the special radius $R=\sqrt{\alpha^{\prime}}$.
These extra states include winding and momentum modes of the string.
To understand the dynamics that result by including these states we can find
the energy density and pressure, which follow from the free
energy as
\begin{eqnarray}
\rho=\frac{1}{V}\frac{\partial}{\partial \beta} (\beta F)
\nonumber \\
p_i=-\frac{1}{9V}\frac{\partial F}{\partial (ln R_i)},
\end{eqnarray}
where $V$ is the spatial volume and $R_i$ is the scale factor in the $ith$
direction.  As discussed above, we take initial conditions where
three dimensions have grown large and six remain near the string
scale.  For such initial data, the above pressure in three
dimensions corresponds to the equation of state of a radiation
dominated universe $p_3=\frac{1}{3} \rho$, whereas the pressure in
the small dimensions gives the behavior,
\begin{equation} \label{eq1}
p_6<0 \;\; \text{for} \;\; R_6>\sqrt{\alpha^{\prime}} \;\;\;\;
\text{and} \;\;\;\; p_6>0 \;\; \text{for} \;\; R_6<\sqrt{\alpha^{\prime}}.
\end{equation}
In {\em dilaton gravity} negative pressure implies a contracting universe, whereas positive pressure
leads to expansion.  Thus, as can be seen in Figure \ref{fig1}, pressure leads to a
stabilizing effect for the scale factor of the extra dimensions
driving the
radius toward the enhanced symmetry point $R_{(6)}=\sqrt{\alpha^{\prime}}$ where the
pressure vanishes.  At this location the gauge symmetry of the
heterotic string is enhanced,
$E_8 \times E_8 + U(1)^6 \longrightarrow E_8 \times E_8 + SU(2)^6$.

In addition to stabilizing the volume moduli, it has been shown that the remnant string
modes, if taken in the dark sector, can also lead to an interesting
cold dark matter candidate\cite{peebles1,peebles2} (see also\cite{thorstenme}).

\section{Moduli Trapping and Stabilization in $4D$}
If one attempts to extend the arguments above to the $4D$
effective field theory, one finds that the stabilization mechanism
no longer holds.  This is not surprising since the pressure in the
extra dimensions has no analog from the $4D$ perspective.
However, one thing that should remain is the idea of enhanced
symmetry.

Recall that it was the contribution of the enhanced symmetry states near
$R=\sqrt{\alpha^{\prime}}$ that led to the pressure terms in
(\ref{eq1}) stabilizing the extra dimensions.  We can account for
these enhanced states from the effective field theory (EFT)
perspective by considering the effects of particle production near the enhanced
symmetry point, $R=\sqrt{\alpha^{\prime}}$.

To understand how this mechanism works let us consider the
simplest case of heterotic strings on the background $\mathbb{R}^4
\times S^1$.  The low energy effective action comes from the compactification of
the action (\ref{action1}).  The dynamics are then given by dilaton
gravity coupled to a chiral $U(1)$ gauge theory,
\begin{equation} \label{l1}
{\mathcal L}_{\text{m}}=(\partial \sigma)^2-\frac{1}{4g^2}(F_{\mu
\nu})^2-\frac{1}{4g^2}(\bar{F}_{\mu
\nu})^2,
\end{equation}
where $F=dA$ ($\bar{F}=d\bar{A}$) is the left (right) gauge theory
resulting from the compactification of the higher dimensional
metric and flux and $g$ is the gauge coupling.  The scalar $\sigma$ gives the radius of the
compactification and can be scaled to measure the departure from the
self-dual radius, i.e. $\sigma=0$ at $R=\sqrt{\alpha^{\prime}}$.

We see that $\sigma$ has only a kinetic term and the lack of a potential
implies the radius is free to take any value.  However, as the
modulus passes near the self-dual radius we have noted that there
are additional massless degrees of freedom.  If our theory is to
be complete these extra degrees of freedom must be included in the
low energy effective action.  This is accomplished by lifting the effective
lagrangian in (\ref{l1}) to a non-abelian gauge theory, in this case chiral $SU(2)$.
We introduce the covariant derivative,
\begin{equation}
D_{\mu}\sigma=\partial_{\mu} \sigma + g A_{\mu} \sigma.
\end{equation}
This leads to a time dependent mass for the new vector $A_{\mu}$
\begin{equation}
\frac{1}{2} g^2 \sigma(t)^2 A_{\mu}^2.
\end{equation}
This time dependent mass implies particle production in our
cosmological space-time.  The stabilization of the radius $\sigma$
can now be realized as follows; initially $\sigma$ is dominated strictly by
the kinetic term, however once it passes
near the enhanced symmetry point $\sigma=0$, $A_{\mu}$ particles
will be produced.  Then, as $\sigma$ continues its trajectory the
mass of the $A_{\mu}$'s will increase and this leads to backreaction on
$\sigma$.  This force, along with friction from the cosmological expansion,
will eventually stabilize $\sigma$ at the enhanced symmetry point $\sigma=0$.

This is a simple example of moduli trapping\cite{kofman,watson,mohaupt}.
Although we have considered
here a simple toy model of a string on a circle, points of
enhanced symmetry are present in nearly all string and M-theory compactifications.
Moreover, it is worth mentioning that this mechanism need not apply
only to volume moduli.  In fact, in Kofman, et. al.\cite{kofman} the
modulus of interest was the distance between two branes which is
of course related by T-duality to the case we have considered here.

\section{Conclusions}
We have seen that from the $10D$ perspective it is possible to
stabilize the volume modulus of a heterotic string
compactification on a $T^6$.  The stabilization was found to be
the result of the pressure exerted on the compact space due to the
presence of enhanced symmetry states contributing to the one-loop
free energy of the strings at finite temperature.  In the $4D$
effective field theory such effects can be understood by
considering particle production near the enhanced
symmetry point.  Near this point additional massless states are allowed which
can be particle produced in the cosmological space-time.  These states then get
masses via the string Higgs Effect and backreact on the modulus stabilizing
the extra dimensions at the self-dual radius.

\section*{Acknowledgments}
I would like to thank Liam McAllister for useful comments and
discussions.  This work was supported by NASA GSRP.

\end{document}